\documentclass[a4paper,11pt]{article}
\pdfoutput=1

\usepackage{pos}
\usepackage{subfig}

\title{Electromagnetic effects on topological observables in QCD}

\author[a]{Bastian B. Brandt}
\author[a]{Gergely Endr\H{o}di}
\author*[a]{Jos\'e Javier Hern\'andez Hern\'andez}
\author[a]{Gergely Mark\'o}
\author[a]{Laurin Pannullo}

\affiliation[a]{Universit{\"a}t Bielefeld,\\
Universit{\"a}tsstra{\ss}e 25, 33615 Bielefeld, Germany}

\emailAdd{brandt@physik.uni-bielefeld.de}
\emailAdd{endrodi@physik.uni-bielefeld.de}
\emailAdd{hernandez@physik.uni-bielefeld.de}
\emailAdd{gmarko@physik.uni-bielefeld.de}
\emailAdd{lpannullo@physik.uni-bielefeld.de}

\abstract{In this proceedings article we present a selected set of our lattice results regarding the effect that background electromagnetic fields have on the topology of QCD. In particular, we report on the lattice spacing-dependence of the axion-photon coupling as well as on the response of the topological susceptibility to strong magnetic fields at nonzero temperatures.  
We demonstrate that the ratio of topological susceptibilities at finite to zero magnetic field has a well behaved continuum limit at low temperatures using a reweighting technique. Moreover, we study the scaling of the axion-photon coupling towards the continuum limit and show that it is less severely affected by discretisation effects.}

\FullConference{The 40th International Symposium on Lattice Field Theory (Lattice 2023)\\
July 31st - August 4th, 2023\\
Fermi National Accelerator Laboratory\\}

\begin{document}
\maketitle

\section{Introduction}

The vacuum of QCD has a non-trivial structure, composed by a combination of different topological sectors, which can be characterised by the topological charge
\begin{equation}
    Q_{\mathrm{top}} = \int d^4x \, q_{\mathrm{top}}, \,\,\,\, q_{\mathrm{top}} = \frac{g^2}{64\pi^2} \epsilon^{\mu\nu\rho\sigma} G_{\mu\nu}^a G_{a,\rho\sigma},
\end{equation}
where $q_{\mathrm{top}}$ is the topological charge density, $g^2$ is the strong coupling constant, $G_{\mu\nu}^a = \partial_{\mu}A^a_{\nu} - \partial_{\nu}A^a_{\mu} -f^{abc}A^b_{\mu}A^c_{\nu}$ is the gluonic field strength, $f^{abc}$ are the structure constants of the $\mathfrak{su}(3)$ algebra and $\epsilon_{\mu\nu\rho\sigma}$ is the Levi-Civita symbol\footnote{We use the convention $\epsilon^{0123}=-1$.}. This CP-odd operator is an integer-valued function on $\mathfrak{su}(3)$ gauge fields, and acts as a label for the corresponding gauge field configuration. The different vacuum topological sectors are distinct in the sense that one cannot change the value of their topological charge by continuous transformations. Moreover the topological charge is invariant under CPT, making it a perfectly valid interaction for QCD (the so-called $\theta$-term). Nevertheless, no CP violation has been observed experimentally. The question of why this operator is highly suppressed is the strong CP problem~\cite{Hook2018tasi}. 

A possible solution to the strong CP problem are axions~\cite{PhysRevLett.38.1440,PhysRevD.16.1791}. Besides providing an explanation for the absence of CP violations in QCD, axions are also dark matter candidates. They are subject to active experimental searches, relying on their coupling to photons. Moreover, their mass (the topological susceptibility $\chi_{\mathrm{top}}$) is a source of cosmological information during the early universe. 
In addition, the impact of magnetic fields on $\chi_{\mathrm{top}}$ (together with temperature effects) is expected to be relevant in off-central heavy-ion collisions, especially in relation to the chiral magnetic effect, which arises due to a combination of magnetic fields and topology~\cite{Fukushima:2008xe}.

This proceedings article is structured as follows: first, we give a brief introduction to the topological susceptibility and the  axion-photon coupling and how can they be computed on the lattice. Then we describe our lattice setup for the simulations and present our preliminary results for the continuum limits of both the axion-photon coupling and the magnetic response of the susceptibility at low temperatures. Finally, we summarise our findings and give a brief outlook.

\section{The topological susceptibility and the axion-photon coupling}

\subsection{The topological susceptibility}

In the absence of a $\theta$-term in the QCD Lagrangian, the expectation value of $Q_{\mathrm{top}}$ vanishes. This leads us to study the size of its fluctuations, which is an inherent property of the vacuum of the theory. These are characterised by the topological susceptibility. It can be defined from the Euclidean partition function of QCD with a $\theta$-term, $Z_{\mathrm{QCD}+\theta}$:
\begin{equation}
    \chi_{\mathrm{top}} =-\frac{1}{\Omega} \left.\frac{\partial^2\ln Z_{\mathrm{QCD}+\theta}}{\partial \theta^2}\right\vert_{\theta = 0}=\frac{\langle Q_{\mathrm{top}}^2\rangle_{\mathrm{QCD}}}{\Omega},
\end{equation}
where $\Omega$ is the space-time volume (an infinite $\Omega$ limit is implicit). We also remark that by promoting $\theta$ to a dynamical field, $\theta\equiv\theta(x)=a(x)/f_a$, where $a(x)$ is the axion field and $f_a$ the axion decay constant\footnote{This is a free parameter in axion models.}, we can associate $\chi_{\mathrm{top}}$ with the mass of the axion: $\chi_{\mathrm{top}}=m_a^2f_a^2$. 

We are interested in the combined temperature and magnetic field dependence of $\chi_{\mathrm{top}}$. The case $T\neq0$, $\Vec{B}=0$, has already been studied perturbatively with chiral perturbation theory (ChPT)~\cite{Hansen:1990yg} and on the lattice~\cite{Bonati2016axion,Borsanyi2016lattice,Jahn:2018dke}. In turn, the impact of magnetic fields on the susceptibility has only been calculated within ChPT. It predicts a mild enhancement of $\chi_{\mathrm{top}}$ with the magnetic field at zero temperature~\cite{Adhikari:2021lbl}. In fact, due to a relation between the topological susceptibility and the chiral condensate that can be proven within leading-order ChPT at zero temperature~\cite{Adhikari:2021lbl}, $\chi_{\mathrm{top}}$ should depend on $\Vec{B}$ in a parametrically similar way as the condensate. This leads to the interesting question, whether a similar relation holds at nonzero temperature as well, i.e.\ whether $\chi_{\mathrm{top}}$ undergoes a type of inverse magnetic catalysis around the QCD transition, just like the condensate~\cite{Bali:2012zg}. In this proceeding we demonstrate that the topological susceptibility is enhanced by the magnetic field at low temperatures and that this behavior is compatible with the ChPT prediction.

\subsection{The axion-photon coupling}
\label{ap_coupling}

The axion couples both directly and indirectly to photons, i.e.\ the coupling is of the form $g_{a\gamma\gamma}=g_{a\gamma\gamma}^0+g_{a\gamma\gamma}^{\mathrm{QCD}}$, where the first term is fixed by the details of the specific axion model (such as KSVZ~\cite{PhysRevLett.43.103} or  DFSZ~\cite{Dine:1981rt}) and the second term depends exclusively on QCD. The QCD part of the coupling has been computed in ChPT to next-to-leading order, giving $g_{a\gamma\gamma}^{\mathrm{QCD}}(\mathrm{NLO})f_a/e^2 =-0.0243(5)$, with $f_a$ being the aforementioned axion decay constant and $e^2$ the electron charge~\cite{di2016QCD}.

By looking at the form of the direct coupling between axions and photons, $g_{a\gamma\gamma}^0a\Vec{E}\cdot\Vec{B}$, we see that the QCD corrections to the coupling can be computed through
\begin{equation}
    g_{a\gamma\gamma}^{\mathrm{QCD}}f_a = \frac{i}{\Omega} \left.\frac{\partial^2 }{\partial \theta\partial(\Vec{E}\cdot\Vec{B})}\ln{Z_{\mathrm{QCD}+\theta+\mathrm{bEM}}}\right\vert_{\theta = \Vec{E} = \Vec{B} = 0},
\end{equation}
where now the partition function includes background electromagnetic fields (bEM) such that $e^2\Vec{E}\cdot\Vec{B}\neq 0$. Note that $\Vec{E}$ denotes the Euclidean (i.e.\ imaginary) electric field. Also, the functional derivative associated to the axion field has been traded with a derivative with respect to $\theta$, since a homogeneous axion field can be interpreted as a $\theta$ parameter.

Hence, we can extract the value of the coupling by taking one or two derivatives of the partition function. This was already investigated in~\cite{Brandt:2022jfk} where it was concluded that the optimal method for this computation is to perform only one derivative with respect to $\theta$, giving us (for weak, static and homogeneous bEM)
\begin{equation}
    \frac{\langle Q_{\mathrm{top}}\rangle_{\Vec{E},\Vec{B}}}{\Omega} = g_{a\gamma\gamma}^{\mathrm{QCD}}f_a \Vec{E}\cdot\Vec{B},
    \label{eq:gagg}
\end{equation}
where $\langle\rangle_{\Vec{E}\cdot\Vec{B}}$ indicates the expectation value in the presence of non-zero bEM. Hence, by studying the response of the topological charge to $\Vec{E}\cdot\Vec{B}$, the QCD corrections to the axion-photon coupling can be extracted via numerical differentiation. Incidentally, a nonzero shift of $\langle Q_{\mathrm{top}}\rangle$ due to $\Vec{E}\cdot\Vec{B}\neq0$ can be explained through the index theorem \cite{Brandt:2022jfk}. We mention that the above coefficient was studied on the lattice in~\cite{DElia:2012ifm}. Moreover, it also plays a role in the context of the chiral magnetic effect in heavy-ion collisions~\cite{Asakawa:2010bu,Bali:2014vja}.

\section{Zero mode reweighting}

Previous lattice calculations have shown that the topological susceptibility suffers from sizeable lattice artifacts that hinder taking a controlled continuum limit~\cite{Borsanyi2016lattice}. The main source of these lattice artifacts is the staggered discretisation used for the fermion fields. On a given gauge configuration with topological charge $Q_{\mathrm{top}}$, we expect the Dirac operator to have $|Q_{\mathrm{top}}|$ exact zero modes in the continuum limit due to the index theorem. However, the staggered Dirac operator lacks these exact zero modes. This means that the weight of a configuration -- the Dirac determinant -- is being overestimated by a factor $\langle W \rangle/W$~\cite{Borsanyi2016lattice}, where
\begin{equation}
    W = \prod_{f}\prod_{i=1}^{2|Q_{\mathrm{top}}|}\left(\frac{m_f^2}{\lambda_i^2+m_f^2}\right)^{n_f/4},
\end{equation}
is what we will call the reweighting factor. Here $n_f$ is the number of quarks of flavour $f$, $m_f$ is the mass of the corresponding flavour and $\lambda_i$ is the $i$-th lowest eigenvalue of the staggered Dirac operator. The exponent of $1/4$ arises due to the rooting procedure and the product in $i$ extends up to the correct number of zero modes, which for the staggered operator is $4|Q_{\rm top}|$ (we also took into account that the eigenvalues come in complex conjugate pairs). By multiplying each configuration by $W/\langle W\rangle$, we can obtain a new estimation of our observable sampled through an improved probability distribution, which is expected to facilitate the continuum limit~\cite{Borsanyi2016lattice}. Hence, we define expectation values of reweighted observables as $\langle \mathcal{O}\rangle_{\mathrm{rw}}\equiv\langle W \mathcal{O}\rangle/\langle W \rangle$.
Since the larger the $|Q_{\rm top}|$, the more the reweighting factor suppresses the corresponding configuration, such a reweighting will reduce the topological susceptibility.
Notice that at low temperatures, the physical interpretation of the above reweighting is complicated by the mixing of would-be zero-modes of topological origin and near-zero-modes building up the chiral condensate.

Regarding the topological susceptibility, we have found that the above reweighting is indispensable to take a controlled continuum limit in the range of lattice spacings that are used in our simulations. We will discuss the details of our simulations below in  Subsec.~\ref{prel_res_topsusc}. A similar reweighting approach can also be applied to the axion-photon coupling. In that case, the main difference is that the number of eigenvalues that one needs to reweight increases with $e^2\Vec{E}\cdot\Vec{B}$, due to the index theorem. This can become computationally challenging depending on the size of the lattice and the respective values of the magnetic and (imaginary) electric fluxes\footnote{On the lattice, $e^2\Vec{E}\cdot\Vec{B}=36\pi^2n_en_b/\Omega$, where $n_e, n_b \in \mathbb{Z}$ and $\Omega = a^4N_s^3 N_t$.}. Our first investigations on small lattices showed furthermore that there is an overlap problem between the simulated distribution and the reweighted one due to the high value of the overall topology. This complicates the implementation of the above described reweighting. However, we observe that -- contrary to the case of $\chi_{\rm top}$ -- a continuum estimate can be provided here without reweighting, too. For details, see Subsec.~\ref{prel_res_coupling}.

\section{Preliminary results}

\subsection{Lattice setup}

For the simulations in this work we have used the following discretisations: the tree-level improved Symanzik action for the gluon fields and the staggered action with stout-improvement for the quarks. The background electromagnetic fields were introduced as $U(1)$ phases multiplying the $SU(3)$ links of QCD. Both the electric and the magnetic field are parallel and point in the positive $z$ direction. We have chosen the electric field to be imaginary in order to avoid the sign problem. The effect of this analytic continuation is trivial for the coupling, being encoded in the behaviour around zero electromagnetic fields. For our fundamental observable, the topological charge, we have selected two different discretisations with $a^2$~\cite{qtop} and $a^4$~\cite{qtop_imp} improvements, which we will refer to as `regular' and `improved', respectively. The simulations were carried out using $2+1$ flavours with physical masses and with electric charges $q_u=2/3$, $q_d=q_s=-1/3$.

In order to renormalise the gluon fields and to obtain close-to-integer values for the discretised topological charge, we have employed the gradient flow technique~\cite{Luscher2010properties}. Since the topological charge must be independent of the flow time $\tau_f$, plateaus as a function of $\tau_f$ are expected. We have always made sure that such plateaus are reached in order to extract the value of the operator.

\subsection{The topological susceptibility}
\label{prel_res_topsusc}

For our study of the topological susceptibility we have generated ensembles for three different magnetic fields, $eB = 0$, $0.5$ and $0.8$ GeV$^2$; in a range of temperatures around the crossover region, $T = 113-300$ MeV. In order to approach the continuum limit, we employed four ensembles with different lattice spacings, of geometry $24^3\times 6$,  $24^3\times 8$, $28^3\times 10$ and $36^3\times 12$. In this proceeding we present results for the susceptibility at zero and intermediate magnetic fields and for the lowest temperature.

\begin{figure}[ht]%
    \centering
    \subfloat{{\includegraphics[scale=0.45]{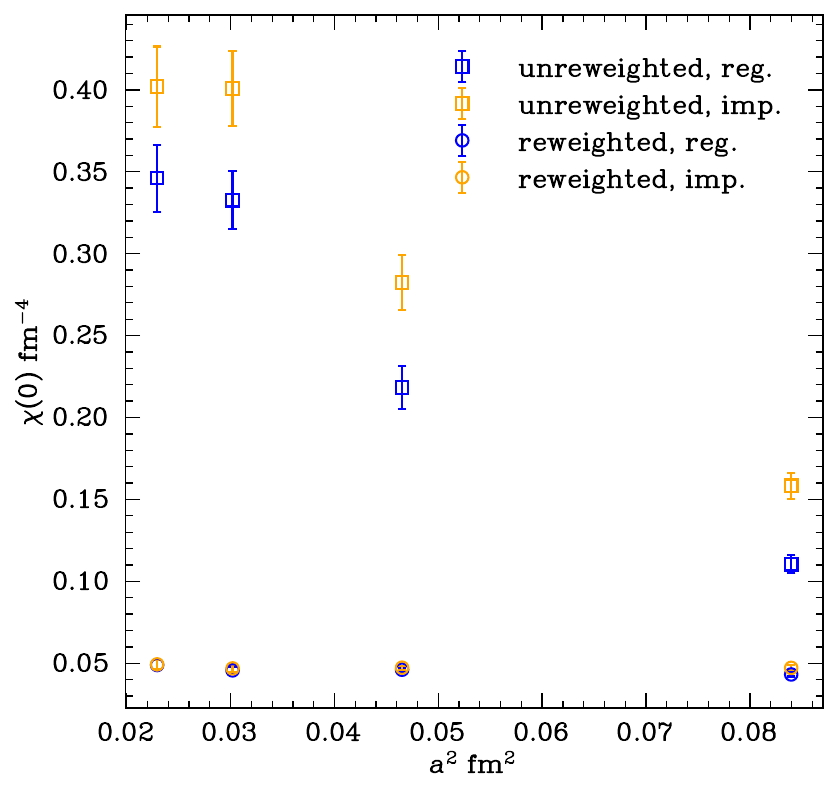} }}%
    \qquad
    \subfloat{{\includegraphics[scale=0.45]{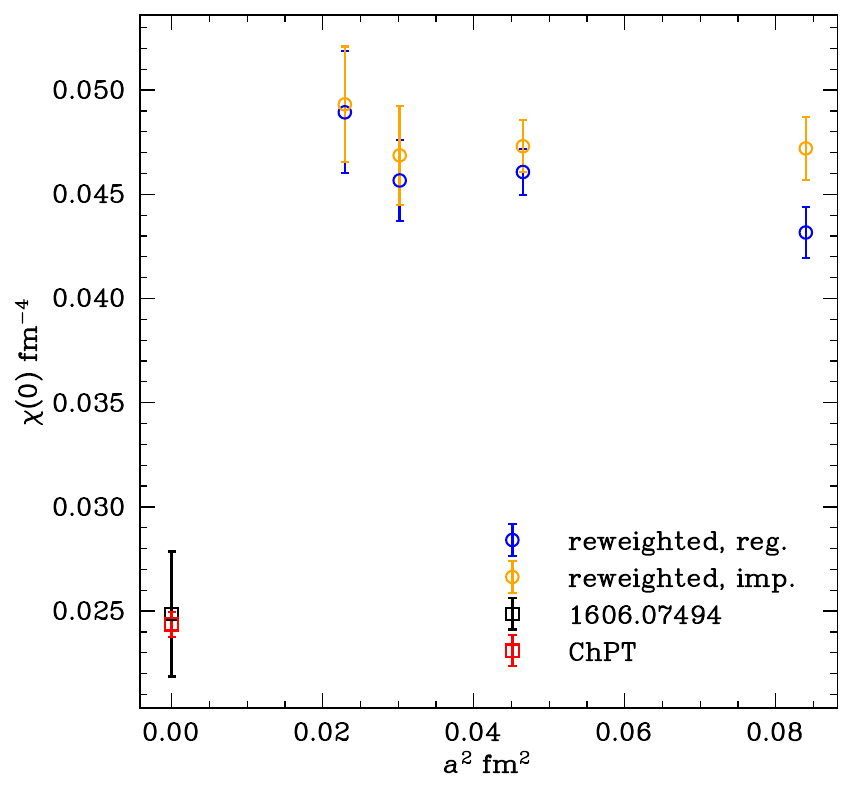} }}%
    \caption{Effect of the reweighting on the topological susceptibility at zero magnetic field for both discretisations of $Q_{\mathrm{top}}$. \textsc{Left panel}: comparison of the reweighted (circles) and unreweighted (squares) susceptibility as a function of the lattice spacing. \textsc{Right panel}: Comparison between the reweighted susceptibility (circles), ChPT~\cite{Adhikari:2021lbl} (red square), and the lattice calculation of Ref.~\cite{Borsanyi2016lattice} (black square).}%
    \label{reweighting_comparison}
\end{figure}

We begin by noticing that at zero magnetic field, a continuum extrapolation of the unreweighted data is not possible, as the results are an order of magnitude larger than the ChPT prediction, see the left panel in Fig.~\ref{reweighting_comparison}. The impact of the reweighting is substantial: it renders the results comparable to the expected continuum value. Still, our lattices are not sufficiently fine to perform a reliable continuum extrapolation, see the right panel of Fig.~\ref{reweighting_comparison}. Here the ChPT prediction~\cite{Adhikari:2021lbl} and the result of previous lattice calculations~\cite{Borsanyi2016lattice} is included as well. 

\begin{figure}[h!]%
    \centering
    \includegraphics[scale=0.45]{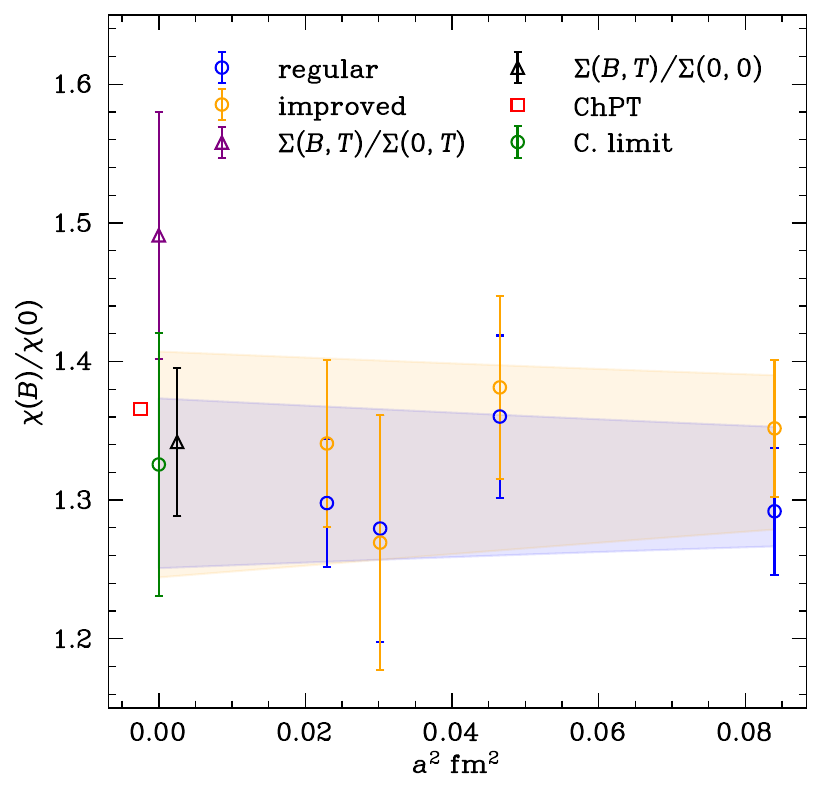}%
    \caption{Ratio of susceptibilities at finite to zero magnetic field for both discretisations of $Q_{\mathrm{top}}$. Open circles correspond to the lattice results, whereas the red square denotes the ChPT prediction~\cite{Adhikari:2021lbl} (displaced to the left for visualisation purposes). The black triangle represents continuum extrapolated lattice results for the ratio of chiral condensates at finite to zero magnetic field~\cite{Bali:2012zg}. For comparison, we include two different combinations of ratios of chiral condensates: one where both numerator and denominator are at the same temperature $T=113$ MeV (purple triangle) and one where the denominator is at $T=0$ (black triangle; displaced to the right for visualisation purposes). The green circle is the result of the fit performed in this study.} 
    \label{ratio_clim}
\end{figure}

Since we are only interested in the magnetic field dependence of $\chi_{\mathrm{top}}$, a strategy to further reduce lattice artifacts is to take the ratio of susceptibilities at finite and zero magnetic field. We find that magnetic field-independent artefacts indeed largely cancel in $\chi_{\rm top} (B)/\chi_{\rm top} (0)$. We present the preliminary result for the continuum limit of the ratio of susceptibilities at $eB = 0.5$ GeV$^2$ and $T = 113$ MeV in Fig.~\ref{ratio_clim}. The errors of the lattice data for the individual susceptibilities contain the error associated to the topological charge not being an exact integer on the lattice added in quadrature. We can observe how the continuum limit is under control for both discretisations of the topological charge. We have fitted the ratios with a linear function in $a^2$ and taken the intercept with zero lattice spacing of the fit on the improved definition as our central value for the continuum result. The difference with the fit for the regular definition has been added in quadrature as a systematic error.
Our result is compared to the ChPT prediction for the ratio of susceptibilities.

Within leading-order ChPT at $T=0$, the magnetic field-dependence of the susceptibility ratio coincides with that of the ratio of chiral condensates $\Sigma(B)/\Sigma(0)$\footnote{$\Sigma (B)$ is the average of the condensates of the $u$ and the $d$ quarks at finite magnetic field.} due to a Gell-Mann-Oakes-Renner type relation~\cite{Adhikari:2021lbl}. To assess the validity of this equivalence, we also compare to the continuum extrapolated lattice results for this ratio~\cite{Bali:2012zg} and indeed find good agreement. 

\subsection{The axion-photon coupling}
\label{prel_res_coupling}

For our study of the axion-photon coupling we have generated ensembles at approximately zero temperature, for three different lattices of geometry $24^3\times32$ (two different lattice spacings), $32^3\times48$ and $40^3\times48$  and with finite background electromagnetic fields, such that $e^2\Vec{E}\cdot\Vec{B}=e^2E_zB_z\neq 0$. We fixed the magnitude of the EM fields approximately within the linear response region of the topological charge. In this proceeding we present the continuum limit for the axion-photon coupling. 

\begin{figure}[ht]%
    \centering
    \includegraphics[scale=0.45]{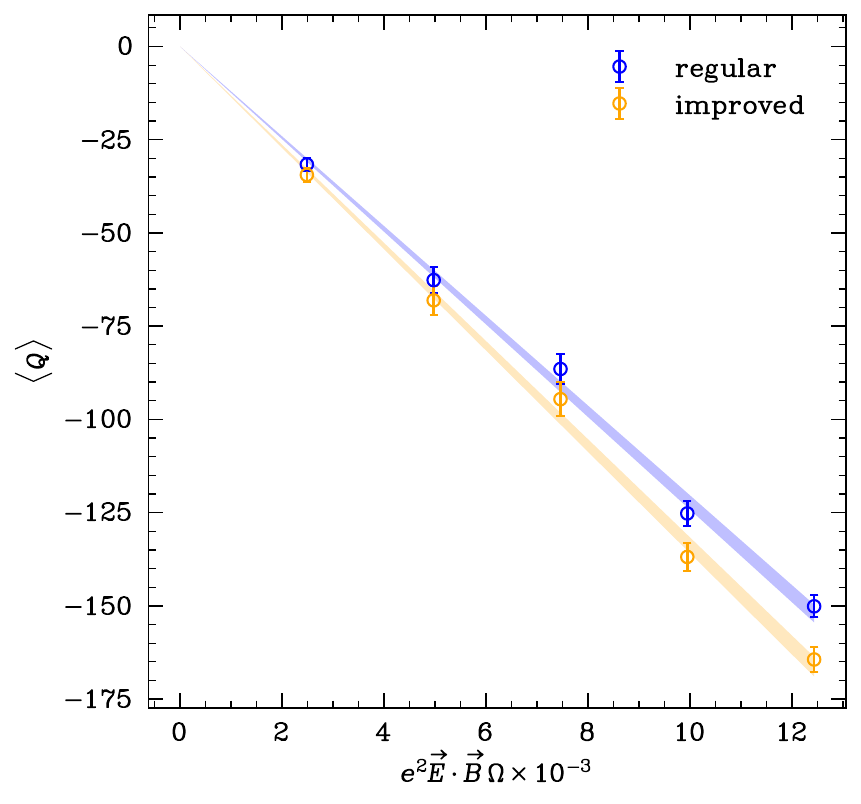}%
    \caption{Expectation value of $Q_{\mathrm{top}}$ as a function of $\Vec{E}\cdot\Vec{B}$ for both discretisations of the topological charge. We observe an approximate linear behaviour in the studied region.}%
    \label{qtop_linear}
\end{figure}

As discussed above, the axion-photon coupling can be determined by studying the dependence of $\langle Q_{\mathrm{top}}\rangle$ on $e^2\Vec{E}\cdot\Vec{B}$ and performing a numerical derivative. In each of our lattices, we observe the approximate linear behaviour~\eqref{eq:gagg} for sufficiently small fields. This is demonstrated for our finest lattice in Fig.~\ref{qtop_linear}.

\begin{figure}[ht]%
    \centering
    \includegraphics[scale=0.45]{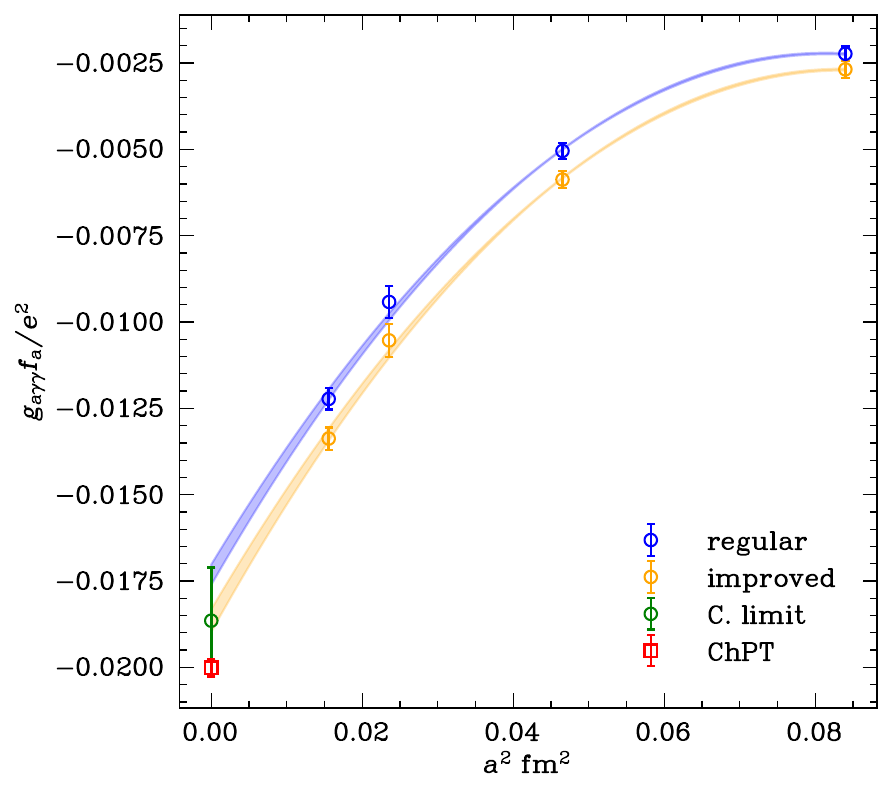}%
    \caption{$g_{a\gamma\gamma}f_a/e^2$ as a function of the lattice spacing for both discretisations of $Q_{\mathrm{top}}$. Open circles correspond to the lattice data, whereas the red square denotes the prediction of ChPT. Our continuum extrapolation (green circle) is compatible with the latter.}%
    \label{coupling_clim}
\end{figure}

We present the preliminary result for the continuum limit of the axion-photon coupling at zero temperature in Fig.~\ref{coupling_clim}. The error of the lattice results contains, besides the statistical error, the uncertainty from the numerical derivative and that associated to the topological charge not being an exact integer. In order to take the continuum limit, we have fitted our lattice points with a quadratic function in $a^2$. We have taken the fit result for the improved definition as our central value and we have added as an error in quadrature the difference with the result of the fit for the regular definition. ChPT provides a prediction for the value of the coupling, which is compatible with our non-perturbative calculation. We notice that, contrary to the case of the susceptibility, the reweighting of the would-be zero modes of the fermion determinant does not seem indispensable for taking the continuum limit. 

Our simulations were performed with degenerate light quarks at the physical point. This is on the basis that isospin effects, associated to the masses of the $u$ and $d$ quarks not being exactly equal, are usually very small. However, certain observables are sensitive to small deviations from this symmetry, like the topological susceptibility. Leading order ChPT calculations show that for the case of symmetric masses, $\chi_{\mathrm{top}}$ is approximately $12\%$ larger than at the asymmetric physical point~\cite{di2016QCD,Borsanyi2016lattice}. The axion-photon coupling faces a similar issue. ChPT predicts that, to leading order, the coupling with non-degenerate masses is larger by a factor~\cite{di2016QCD} 
\begin{equation}
    \frac{2}{5}\frac{m_u+4m_d}{m_u+m_d} \approx 1.21.
\end{equation}
In order to include the effect of asymmetric masses into our analysis, we rescale the value of the coupling obtained through our simulations with degenerate masses by the above factor. Hence, we quote that the preliminary result for the first non-perturbative calculation in full QCD of the model independent part of the axion-photon coupling at the physical point is
\begin{equation}
    g_{a\gamma\gamma}^{\mathrm{QCD}}f_a/e^2 = -0.023(2).
\end{equation}

\section{Summary and further work}

In this proceeding we have investigated the impact of background electromagnetic fields on topological observables in QCD. In particular, we discussed the continuum extrapolations of two specific observables: the ratio of topological susceptibilities at nonzero to zero magnetic field, as well as the axion-photon coupling. We have demonstrated that the topological susceptibility is enhanced by the magnetic field at low temperatures, compatible with the prediction of ChPT. Finally, we have presented the results of the first non-perturbative calculation of the QCD corrections to the axion-photon coupling at the physical point.

We are currently working towards increasing the precision in our computation of $g_{a\gamma\gamma}^{\mathrm{QCD}}$ and in taking the continuum limit for the ratio of susceptibilities at higher magnetic fields and higher temperatures.

\paragraph{Acknowledgments}

This work was supported by the Deutsche Forschungsgemeinschaft (DFG, German Research Foundation) – project number 315477589 – TRR 211 and by the Helmholtz Graduate School for Hadron and Ion Research (HGS-HIRe for FAIR). The authors are also grateful to Guy D. Moore for fruitful discussions. The computations in this work were performed on the GPU cluster at Bielefeld University.
Some of the calculations were carried out using the \texttt{SIMULATeQCD} framework \cite{Mazur:2021zgi,HotQCD:2023ghu}. 

\bibliographystyle{utphys} 
\bibliography{skeleton}

\providecommand{\href}[2]{#2}\begingroup\raggedright\begin{thebibliography}{10}

\bibitem{Hook2018tasi}
A.~Hook {\em PoS} {\bfseries TASI2018} (2019) 004,
  \href{https://arxiv.org/abs/1812.02669}{{\ttfamily arXiv:1812.02669
  [hep-ph]}}.

\bibitem{PhysRevLett.38.1440}
R.~D. Peccei and H.~R. Quinn
  \href{https://dx.doi.org/10.1103/PhysRevLett.38.1440}{{\em Phys. Rev. Lett.}
  {\bfseries 38} (Jun, 1977) 1440--1443}.

\bibitem{PhysRevD.16.1791}
R.~D. Peccei and H.~R. Quinn
  \href{https://dx.doi.org/10.1103/PhysRevD.16.1791}{{\em Phys. Rev. D}
  {\bfseries 16} (Sep, 1977) 1791--1797}.

\bibitem{Fukushima:2008xe}
K.~Fukushima, D.~E. Kharzeev, and H.~J. Warringa
  \href{https://dx.doi.org/10.1103/PhysRevD.78.074033}{{\em Phys. Rev. D}
  {\bfseries 78} (2008) 074033},
  \href{https://arxiv.org/abs/0808.3382}{{\ttfamily arXiv:0808.3382 [hep-ph]}}.

\bibitem{Hansen:1990yg}
F.~C. Hansen and H.~Leutwyler
  \href{https://dx.doi.org/10.1016/0550-3213(91)90259-Z}{{\em Nucl. Phys. B}
  {\bfseries 350} (1991) 201--227}.

\bibitem{Bonati2016axion}
C.~Bonati, M.~D'Elia, M.~Mariti, G.~Martinelli, M.~Mesiti, F.~Negro,
  F.~Sanfilippo, and G.~Villadoro
  \href{https://dx.doi.org/10.1007/JHEP03(2016)155}{{\em JHEP} {\bfseries 03}
  (2016) 155}, \href{https://arxiv.org/abs/1512.06746}{{\ttfamily
  arXiv:1512.06746 [hep-lat]}}.

\bibitem{Borsanyi2016lattice}
S.~Bors\'anyi {\em et~al.} \href{https://dx.doi.org/10.1038/nature20115}{{\em
  Nature} {\bfseries 539} no.~7627, (2016) 69--71},
  \href{https://arxiv.org/abs/1606.07494}{{\ttfamily arXiv:1606.07494
  [hep-lat]}}.

\bibitem{Jahn:2018dke}
P.~T. Jahn, G.~D. Moore, and D.~Robaina
  \href{https://dx.doi.org/10.1103/PhysRevD.98.054512}{{\em Phys. Rev. D}
  {\bfseries 98} no.~5, (2018) 054512},
  \href{https://arxiv.org/abs/1806.01162}{{\ttfamily arXiv:1806.01162
  [hep-lat]}}.

\bibitem{Adhikari:2021lbl}
P.~Adhikari \href{https://dx.doi.org/10.1016/j.physletb.2021.136826}{{\em Phys.
  Lett. B} {\bfseries 825} (2022) 136826},
  \href{https://arxiv.org/abs/2103.05048}{{\ttfamily arXiv:2103.05048
  [hep-ph]}}.

\bibitem{Bali:2012zg}
G.~S. Bali, F.~Bruckmann, G.~Endr\H{o}di, Z.~Fodor, S.~D. Katz, and
  A.~Sch{\"a}fer \href{https://dx.doi.org/10.1103/PhysRevD.86.071502}{{\em
  Phys. Rev. D} {\bfseries 86} (2012) 071502},
  \href{https://arxiv.org/abs/1206.4205}{{\ttfamily arXiv:1206.4205
  [hep-lat]}}.

\bibitem{PhysRevLett.43.103}
J.~E. Kim \href{https://dx.doi.org/10.1103/PhysRevLett.43.103}{{\em Phys. Rev.
  Lett.} {\bfseries 43} (Jul, 1979) 103--107}.

\bibitem{Dine:1981rt}
M.~Dine, W.~Fischler, and M.~Srednicki
  \href{https://dx.doi.org/10.1016/0370-2693(81)90590-6}{{\em Phys. Lett. B}
  {\bfseries 104} (1981) 199--202}.

\bibitem{di2016QCD}
G.~Grilli~di Cortona, E.~Hardy, J.~Pardo~Vega, and G.~Villadoro
  \href{https://dx.doi.org/10.1007/JHEP01(2016)034}{{\em JHEP} {\bfseries 01}
  (2016) 034}, \href{https://arxiv.org/abs/1511.02867}{{\ttfamily
  arXiv:1511.02867 [hep-ph]}}.

\bibitem{Brandt:2022jfk}
B.~Brandt, F.~Cuteri, G.~Endr\H{o}di, J.~J.~H. Hern\'andez, and G.~Mark\'o
  \href{https://dx.doi.org/10.22323/1.430.0174}{{\em PoS} {\bfseries
  LATTICE2022} (2023) 174}, \href{https://arxiv.org/abs/2212.03385}{{\ttfamily
  arXiv:2212.03385 [hep-lat]}}.

\bibitem{DElia:2012ifm}
M.~D'Elia, M.~Mariti, and F.~Negro
  \href{https://dx.doi.org/10.1103/PhysRevLett.110.082002}{{\em Phys. Rev.
  Lett.} {\bfseries 110} no.~8, (2013) 082002},
  \href{https://arxiv.org/abs/1209.0722}{{\ttfamily arXiv:1209.0722
  [hep-lat]}}.

\bibitem{Asakawa:2010bu}
M.~Asakawa, A.~Majumder, and B.~Muller
  \href{https://dx.doi.org/10.1103/PhysRevC.81.064912}{{\em Phys. Rev. C}
  {\bfseries 81} (2010) 064912},
  \href{https://arxiv.org/abs/1003.2436}{{\ttfamily arXiv:1003.2436 [hep-ph]}}.

\bibitem{Bali:2014vja}
G.~S. Bali, F.~Bruckmann, G.~Endr\H{o}di, Z.~Fodor, S.~D. Katz, and
  A.~Sch{\"a}fer \href{https://dx.doi.org/10.1007/JHEP04(2014)129}{{\em JHEP}
  {\bfseries 04} (2014) 129}, \href{https://arxiv.org/abs/1401.4141}{{\ttfamily
  arXiv:1401.4141 [hep-lat]}}.

\bibitem{qtop}
S.~Bors\'anyi {\em et~al.}
  \href{https://dx.doi.org/10.1007/JHEP09(2012)010}{{\em JHEP} {\bfseries 09}
  (2012) 010}, \href{https://arxiv.org/abs/1203.4469}{{\ttfamily
  arXiv:1203.4469 [hep-lat]}}.

\bibitem{qtop_imp}
S.~O. Bilson-Thompson, D.~B. Leinweber, and A.~G. Williams
  \href{https://dx.doi.org/10.1016/S0003-4916(03)00009-5}{{\em Annals Phys.}
  {\bfseries 304} (2003) 1--21},
  \href{https://arxiv.org/abs/hep-lat/0203008}{{\ttfamily
  arXiv:hep-lat/0203008}}.

\bibitem{Luscher2010properties}
M.~L\"uscher \href{https://dx.doi.org/10.1007/JHEP08(2010)071}{{\em JHEP}
  {\bfseries 08} (2010) 071}, \href{https://arxiv.org/abs/1006.4518}{{\ttfamily
  arXiv:1006.4518 [hep-lat]}}. [Erratum: JHEP 03, 092 (2014)].

\bibitem{Mazur:2021zgi}
L.~Mazur, \href{https://dx.doi.org/10.4119/unibi/2956493}{{\em {Topological
  Aspects in Lattice QCD}}}.
\newblock PhD thesis, Bielefeld U., 2021.

\bibitem{HotQCD:2023ghu}
{\bfseries HotQCD} Collaboration, L.~Mazur {\em et~al.}
  \href{https://arxiv.org/abs/2306.01098}{{\ttfamily arXiv:2306.01098
  [hep-lat]}}.

\end{thebibliography}\endgroup

\end{document}